\documentclass[nofootinbib,showpacs,superscriptaddress,11pt]{revtex4-1}

\usepackage{graphicx,enumitem,natbib}
\usepackage{bm,mathrsfs}

\begin{document}

\def\imag{{\rm i}\,\!}

\def\unit#1{\mathop{\mathbf{1}_#1}}
\def\op#1#2{\mathop{\mathscr{#1} \setbox0\hbox{$#2$}
	\ifnum\wd0=0\else\kern -1pt\left(#2\right)\fi}}

\title{On the Instrument Profile of Slit Spectrographs}

\author{R.\ Casini}
\author{A.\ G.\ de Wijn}

\affiliation{High Altitude Observatory, National Center for Atmospheric
Research,\footnote{The National Center for Atmospheric Research is
sponsored by the National Science Foundation}\break P.O.\ Box 3000, Boulder, 
CO 80307-3000}

\begin{abstract}
We derive an analytic expression for the instrument profile of a slit 
spectrograph, also known as the line spread function. While this problem 
is not new, our treatment relies on the operatorial approach to the 
description of diffractive optical systems, which provides a general 
framework for the analysis of the performance of slit spectrographs 
under different illumination conditions. 
Based on our results, we propose an approximation to the spectral 
resolution of slit spectrographs, taking into account diffraction
effects and sampling by the detector, which improves upon the often 
adopted approximation based on the root-sum-square of the individual 
contributions from the slit, the grating, and the detector pixel.
\end{abstract}


\maketitle

\section{Introduction}
\label{sec:intro}

The quantitative analysis 
of spectroscopic data from observations demands knowledge of the 
instrument profile of the spectrographic tools used 
for data acquisition. The instrument profile determines the spectral 
resolution of the observations, which ultimately affects the sensitivity 
of measurements of physically relevant plasma quantities, such as line 
widths (e.g., radiative, collisional, thermal, plasma turbulence), 
line shifts (e.g., isotopic, Doppler, gravitational), and spectral 
modulations in both intensity and polarization (e.g., Zeeman and Stark
splitting by applied fields, spectral line polarization in anisotropic 
media, redistribution of radiation frequency in partially coherent
scattering).

In particular, the instrument profile of a slit-based spectrograph, 
also called the \emph{line spread function} (LSF), depends on the aperture 
of the entrance slit, the \emph{finesse} of the spectral profile of 
the dispersive element (e.g., prism, diffraction grating), and the 
sampling length of the imaged spectrum at the detector (e.g., the 
pixel size of a CCD camera, or the average size of film grains).

Often, for computational convenience, the simplistic assumption 
is made that the width of the instrument profile (typically, its 
\emph{full width at half maximum}, FWHM) can be calculated as the
\emph{root-sum-square} (RSS) of the individual contributions from the slit, the
chromatic dispersion profile, and the detector's sampling 
(see, e.g., \cite{LT98,El08}). 
This approximation is of course based on the assumption that the profiles 
associated with each of these contributions can be represented
by Gaussian distributions, and that there are no inter-dependences 
among those contributions.
While this assumption may turn out to be valid in some particular 
configurations of a slit-based spectrograph, it is not
representative of the general dispersive and imaging properties of 
a given instrument. 
In particular, this approximation cannot properly account for the 
different diffraction conditions that the experimenter encounters when 
exploring a very large interval of wavelengths -- say, from the extreme 
blue to the near infrared -- while using a fixed aperture of the slit. 
It also fails to reproduce the different sampling regimes at the 
detector that necessarily occur in such multi-wavelength usage of the
spectrograph, and which may not correspond exactly to the desired 
sampling condition (e.g., pixel matching, Nyquist sampling).

\begin{figure}[t!]
\centering
\includegraphics[width=.8\hsize]{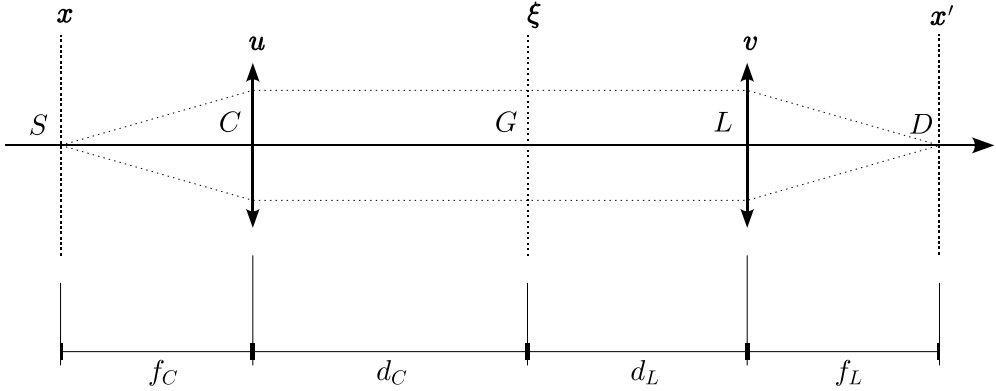}\vspace{-2pt}
\caption{\label{fig:layout}
Layout of the imaging system. See text for details.}
\end{figure}

In this note, we investigate the diffraction properties of a 
spectrograph in order to arrive at a 
general expression for the LSF, and ultimately to an analytic
approximation to the spectral resolution of the instrument. 
Obviously, we must rely on the Fourier analysis of the instrument's 
components. The problem is not new, and in fact we re-derive known
results that date at least as far back as Wadsworth's extensive 
investigation of the effects of the slit aperture on the shape of 
spectral lines \cite{Wa97} (see also \cite{VC30,Mi67} and references 
therein). However, for our treatment of the problem, we adopted the 
tools of the 
operational calculus of Fourier optics (e.g., \cite{NS80,Go96}, 
and references therein), which lead to a very general and compact 
description of the \emph{optical transfer operator} (OTO) of a slit 
spectrograph, opening the ground for further investigations of the 
performance of such instruments under different illumination conditions, 
through both analytic and numerical approaches.

In Sect.~\ref{sec:derivation} we derive the general LSF of a slit 
spectrograph under the hypothesis of coherent illumination, and obtain 
a closed analytic form for the case where the grating defines 
the limiting aperture of the system. We then show how this approach
immediately leads to the expression of the LSF in the case of
incoherent illumination. Finally, in Sect.~\ref{sec:resolution} we propose 
a functional form for the FWHM of the instrument profile after sampling 
by the detector, for the purpose of estimating the spectral 
resolution of the instrument.

\section{Derivation of the Line Spread Function}
\label{sec:derivation}

We consider a slit spectrograph (see Fig.~\ref{fig:layout}), 
consisting of a slit ($S$), a collimator lens ($C$), a grating ($G$), 
a camera lens ($L$), and a detector ($D$).
We assume that the spectrograph can be described operationally as a 
linear system $\mathscr{S}$ acting on the input field $U$ at the slit, 
and producing an output field $U'$ at the detector, so that 
$U'=\mathscr{S} U$.
The OTO for such a system, assumed to be perfectly stigmatic
(i.e., free of optical aberrations), takes the form (see \cite{Go96}, 
and the Appendix)
\begin{eqnarray} \label{eq:OTO}
\op{S}{}
&=&\op{R}{f_L} \op{Q}{-\frac{1}{f_L}} \unit{L} \op{R}{d_L} 
	\unit{G} 
\op{R}{d_C} \unit{C} \op{Q}{-\frac{1}{f_C}} \op{R}{f_C}
	\unit{S}\;.
\end{eqnarray}
Here the operator $\op{Q}{-1/f}$ describes the wavefront modification
introduced by a powered optic of focal length $f$, whereas $\op{R}{d}$ 
describes the free-space propagation of the wavefront along the 
optical path of length $d$ (see the Appendix).
Equation~(\ref{eq:OTO}) does not take into account the presence of
beam-limiting apertures along the optical path. The location of those
apertures is marked by the presence of the unit operator. 

We want to determine the instrument profile (or LSF) for a given 
wavelength, $\lambda$. Using the transformation rules of the operational 
calculus of Fourier optics (see Appendix), and introducing the
dimensional constant $k=1/(\lambda^2\,f_L f_C)$, we find, apart from a 
constant phase factor,
\begin{eqnarray} \label{eq:genOTO}
\op{S}{}
&=&k \left\{
	\op{Q}{\frac{1}{f_L}} \op{V}{\frac{1}{\lambda f_L}} \mathscr{F}
	\op{Q}{\frac{1}{f_L}} 
\right\}
	\op{Q}{-\frac{1}{f_L}} \unit{L}
\left\{ 
	\mathscr{F}^{-1} \op{Q}{-\lambda^2 d_L} \mathscr{F}
\right\} \unit{G} \nonumber \\
&&\kern 1cm
	\times \left\{ 
	\mathscr{F}^{-1} \op{Q}{-\lambda^2 d_C} \mathscr{F}
\right\} \unit{C}
	\op{Q}{-\frac{1}{f_C}}
\left\{
	\op{Q}{\frac{1}{f_C}} \op{V}{\frac{1}{\lambda f_C}} \mathscr{F}
	\op{Q}{\frac{1}{f_C}}
\right\} \unit{S} \nonumber \\
&=&k \op{Q}{\frac{1}{f_L}} \op{V}{\frac{1}{\lambda f_L}} \mathscr{F}\,\unit{L}
\left\{ 
	\mathscr{F}^{-1} \op{Q}{-\lambda^2 d_L} \mathscr{F}
\right\} \unit{G} \nonumber \\
&&\kern 1cm
	\times \left\{ 
	\mathscr{F}^{-1} \op{Q}{-\lambda^2 d_C} \mathscr{F}
\right\} \unit{C}
	\op{V}{\frac{1}{\lambda f_C}} \mathscr{F}
	\op{Q}{\frac{1}{f_C}} \unit{S}\;,
\end{eqnarray}
where $\mathscr{F}$ represents the operation of Fourier transform, and
$\op{V}{c}$ is the scaling operator by the quantity $c$ (see Appendix).


In order to develop the OTO expression beyond the general form 
of Eq.~(2), we need to make some assumptions on the transmission 
functions of the various apertures. First of all, we will assume that 
the camera lens, $L$, is sized 
such as not to introduce any vignetting of the beam after the collimator 
lens and the grating. In such cases, we can assume that the diameter of 
the camera lens is virtually infinite, and 
noting that $\mathscr{F}\unit{L}\mathscr{F}^{-1}=1$,
we can rewrite, 
\begin{eqnarray}
\mathscr{S}
&=&k \op{Q}{\frac{1}{f_L}} \op{V}{\frac{1}{\lambda f_L}}
	\op{Q}{-\lambda^2 d_L} \mathscr{F} \unit{G} \nonumber \\
&&\kern 1cm
	\times \left\{ 
	\mathscr{F}^{-1} \op{Q}{-\lambda^2 d_C} \mathscr{F}
\right\} \unit{C}
	\op{V}{\frac{1}{\lambda f_C}} \mathscr{F}
	\op{Q}{\frac{1}{f_C}} \unit{S}\;.
\end{eqnarray}
In fact, because the beam between $C$ and $L$ is collimated, we can
safely assume that the vignetting of the beam is produced only by the
smallest size aperture in the collimator. This can be either the
collimator lens, or the grating surface. In the first case, Eq.~(3)
becomes, 
using $\mathscr{F}\unit{G}\mathscr{F}^{-1}=1$,
\begin{eqnarray} \label{eq:eq4a}
\mathscr{S}_C
&=&k \op{Q}{\frac{1}{f_L}} \op{V}{\frac{1}{\lambda f_L}}
	\op{Q}{-\lambda^2 [d_L+d_C]} \mathscr{F} \unit{C}
	\op{V}{\frac{1}{\lambda f_C}} \mathscr{F}
	\op{Q}{\frac{1}{f_C}} \unit{S} \nonumber \\
&=&k \op{Q}{\frac{1}{f_L}} \op{Q}{-\frac{d_L+d_C}{f_L^2}} 
	\op{V}{\frac{1}{\lambda f_L}} \mathscr{F} \unit{C}
	\op{V}{\frac{1}{\lambda f_C}} \mathscr{F}
	\op{Q}{\frac{1}{f_C}} \unit{S} \nonumber \\
&=&k \op{Q}{\frac{1}{f_L^2}\!\left[f_L-d_L-d_C\right]} 
	\op{V}{\frac{1}{\lambda f_L}} \mathscr{F} \unit{C}
	\op{V}{\frac{1}{\lambda f_C}} \mathscr{F}
	\op{Q}{\frac{1}{f_C}} \unit{S}\;,
\end{eqnarray}
where in the second line we used the commutation property of $\op{Q}{}$
and $\op{V}{}$, and in the third line the group property of $\op{Q}{}$.
This expression shows that the field reaching the collimator lens is a
scaled Fourier transform (FT) of the input field multiplied by the slit
transmission function and an exponential phase factor determined by the 
focal length $f_C$. Then the signal that is transferred down to the
detector is another scaled FT, followed by another phase factor, which,
however, is inessential for the determination of the LSF of the
spectrograph.

The alternate approach is to assume that the only limiting aperture is the 
one imposed by the finite size of the grating. In such case we can 
ignore the aperture of the collimator lens, but the manipulation of Eq.~(3) 
is somewhat more complicated than in the case leading to
Eq.~(\ref{eq:eq4a}).
We make use of the following relations,
\begin{eqnarray}
\mathscr{F}^{-1}
&=&\op{V}{-1}\mathscr{F}=\mathscr{F}\op{V}{-1}\;, \\
\mathscr{F}\op{V}{\frac{1}{\lambda f_C}}\mathscr{F}
&=&\mathscr{F}^2\op{V}{\lambda f_C}\equiv\op{V}{-1}\op{V}{\lambda f_C}
	=\op{V}{-\lambda f_C}\;,
\end{eqnarray}
through which Eq.~(3) transforms into
\begin{eqnarray}
\mathscr{S}_G \label{eq:eq4b}
&=&k \op{Q}{\frac{1}{f_L}} \op{V}{\frac{1}{\lambda f_L}}
	\op{Q}{-\lambda^2 d_L} \mathscr{F} \unit{G} \mathscr{F}
	\op{V}{-1}
	\op{Q}{-\lambda^2 d_C} \op{V}{-\lambda f_C}
	\op{Q}{\frac{1}{f_C}} \unit{S} \nonumber \\
&=&k \op{Q}{\frac{1}{f_L}} \op{Q}{-\frac{d_L}{f_L^2}} 
	\op{V}{\frac{1}{\lambda f_L}}
	\mathscr{F} \unit{G} \mathscr{F}
	\op{V}{-1}
	\op{V}{-\lambda f_C}
	\op{Q}{-\frac{d_C}{f_C^2}} 
	\op{Q}{\frac{1}{f_C}} \unit{S} \nonumber \\
&=&k \op{Q}{\frac{1}{f_L}\!\left[1-\frac{d_L}{f_L}\right]} 
	\op{V}{\frac{1}{\lambda f_L}}
	\mathscr{F} \unit{G} \mathscr{F}
	\op{V}{\lambda f_C}
	\op{Q}{\frac{1}{f_C}\!\left[1-\frac{d_C}{f_C}\right]} \unit{S}
\nonumber \\
&=&k \op{Q}{\frac{1}{f_L}\!\left[1-\frac{d_L}{f_L}\right]} 
	\op{V}{\frac{1}{\lambda f_L}}
	\mathscr{F} \unit{G} 
	\op{V}{\frac{1}{\lambda f_C}}
	\mathscr{F}
	\op{Q}{\frac{1}{f_C}\!\left[1-\frac{d_C}{f_C}\right]} \unit{S}\;. 
\end{eqnarray}

The optical configuration described by this last expression has an
advantage for the determination of the spectrograph's instrument
profile. We note that, just like in the case of Eq.~(\ref{eq:eq4a}), the
transfer of the field amplitude from the slit plane to the vignetting
aperture is again a scaled FT of the input field multiplied by the slit
transmission function and by an exponential phase factor. However, in 
this case, the distance 
of the grating from the collimator lens can be chosen so that $d_C=f_C$,
in which case the phase factor is equal to 1. Then the system can be
described operationally as a scaled FT of the illuminated slit aperture, 
which is then multiplied by the grating transfer function, and then 
further FT-ed and magnified onto the detector. The final phase 
factor again is inessential for the determination of the 
spectrograph's LSF. In fact, this phase factor could effectively be 
reduced to 1 as well, although the necessary condition $d_L=f_L$ may be 
hard to meet in realistic spectrograph designs, because $f_L$ must be 
subject to the constraint of the spatial/spectral sampling requirement, 
and thus it is driven by the pixel size of the detector.

For the following development, we must introduce explicitly the 
transmission functions of the various apertures.
If we use Eq.~(\ref{eq:eq4a}) for the computation of the spectrograph's 
LSF, and ignore the final phase factor, we then have
\begin{eqnarray}
\mathscr{S}_C\,U
&=&k \op{V}{\frac{1}{\lambda f_L}} \mathscr{F} t_C\,
	\op{V}{\frac{1}{\lambda f_C}} \mathscr{F} \op{Q}{\frac{1}{f_C}}
	t_S\,U \nonumber \\
&=&k \op{V}{\frac{1}{\lambda f_L}} \mathscr{F} t_C\,
	\mathscr{F} \op{V}{\lambda f_C} \op{Q}{\frac{1}{f_C}}
	t_S\,U \nonumber \\
&\equiv& k \op{V}{\frac{1}{\lambda f_L}} \mathscr{F} t_C\,p
	= k \op{V}{\frac{1}{\lambda f_L}} \left(
	\tilde{t}_C \ast \tilde{p} \right)\;,
\end{eqnarray}
where in the last line we defined
$p\equiv\mathscr{F} \op{V}{\lambda f_C} 
	\op{Q}{1/f_C} t_S\,U$, 
and used the convolution theorem
(the symbol ``$\,\widetilde{\hphantom{1}}\,$'' indicates the operation 
of FT, and ``$\ast$'' is the
convolution product). Using the rules given in the Appendix,
\begin{equation}
\tilde{p}
=\mathscr{F}^2 \op{V}{\lambda f_C} 
	\op{Q}{1/f_C} t_S\,U \\
\equiv \op{V}{-\lambda f_C} 
	\op{Q}{1/f_C} t_S\,U\;,
\end{equation}
so that (cf.~Eq.~[\ref{eq:scaling2}])
\begin{eqnarray} \label{eq:eq5a}
\mathscr{S}_C\,U
&=&k \op{V}{\frac{1}{\lambda f_L}} \tilde{t}_C \ast 
	\op{V}{\frac{1}{\lambda f_L}} \tilde{p} \nonumber \\ 
&=&k \left(\tilde{t}_C\circ\left[\frac{1}{\lambda f_L}\right]\right) \ast 
	\op{V}{-\frac{f_C}{f_L}} \op{Q}{\frac{1}{f_C}}
	t_S\,U \nonumber \\
&=&k \left(\tilde{t}_C\circ\left[\frac{1}{\lambda f_L}\right]\right) \ast 
	\op{Q}{\frac{f_C}{f_L^2}} \op{V}{-\frac{f_C}{f_L}}
	t_S\,U \nonumber \\
&=&k \left(\tilde{t}_C\circ\left[\frac{1}{\lambda f_L}\right]\right) \ast 
	\op{Q}{\frac{f_C}{f_L^2}} 
	\left(t_S\circ\left[-\frac{f_C}{f_L}\right]\right)
	\left(U\circ\left[-\frac{f_C}{f_L}\right]\right)\;.
\end{eqnarray}
where we indicated with ``$\circ$'' the operation of function
composition.

Using instead Eq.~(\ref{eq:eq4b}) with $d_C=f_C$, and again ignoring 
the final phase factor, we find, with a manipulation similar to the 
previous one,
\begin{eqnarray} \label{eq:eq5b}
\mathscr{S}_G\,U
&=&k \op{V}{\frac{1}{\lambda f_L}} \mathscr{F} t_G
	\op{V}{\frac{1}{\lambda f_C}} \mathscr{F} 
	t_S\,U 
=\cdots \nonumber \\
&=&k \left(\tilde{t}_G\circ\left[\frac{1}{\lambda f_L}\right]\right) \ast 
	\left(t_S\circ\left[-\frac{f_C}{f_L}\right]\right)
	\left(U\circ\left[-\frac{f_C}{f_L}\right]\right)\;.
\end{eqnarray}

Hence, the field amplitude at the detector, for the two cases
where the limiting aperture is provided, respectively, by the collimator or 
the grating, becomes 
\begin{eqnarray}
U_C(\bm{x'}) \label{eq:eq6a}
&=&k \int d\bm{\chi}\;
	\tilde{t}_C\biggl(\frac{1}{\lambda f_L}
	\left(\bm{x'}-\bm{\chi}\right)\biggr) 
	\exp\biggl(\imag\,\frac{\pi}{\lambda}\,\frac{f_C}{f_L^2}\,
		|\bm{\chi}|^2\biggr)\, 
	t_S\biggl(-\frac{f_C}{f_L}\bm{\chi}\biggr)
	U\biggl(-\frac{f_C}{f_L}\bm{\chi}\biggr)\;, \\
U_G(\bm{x'}) \label{eq:eq6b}
&=&k \int d\bm{\chi}\;
	\tilde{t}_G\biggl(\frac{1}{\lambda f_L}
	\left(\bm{x'}-\bm{\chi}\right)\biggr)\,
	t_S\biggl(-\frac{f_C}{f_L}\bm{\chi}\biggr)
	U\biggl(-\frac{f_C}{f_L}\bm{\chi}\biggr)\;.
\end{eqnarray}

For the purpose of this work, we consider the simplest case of a 
fully coherent, input radiation field of unit amplitude, i.e., $U(\bm{x})=1$.
Then, in the case of Eq.~(\ref{eq:eq6b}), the field amplitude at 
the detector is given by the convolution of the slit transfer 
function (a box)---projected onto the detector, and properly scaled by 
the spectrograph magnification---with the FT of the grating aperture (a 
``sinc'' function; see \cite{Go96}). That expression can be formally 
integrated in terms of the ``sine integral'' function,
$\hbox{Si}(\theta)$,
%
\begin{eqnarray} \label{eq:field}
U_G(x',y')=\frac{1}{\pi^2}
&&\left[
	\hbox{Si}\left(
	\frac{\pi}{2}\,\frac{w_S/f_C-2x'/f_L}{\lambda/w_G} 
	\right) +
	\hbox{Si}\left(
	\frac{\pi}{2}\,\frac{w_S/f_C+2x'/f_L}{\lambda/w_G} 
	\right)
	\right] \nonumber \\
\times
&&\left[
	\hbox{Si}\left(
	\frac{\pi}{2}\,\frac{h_S/f_C-2y'/f_L}{\lambda/h_G} 
	\right) +
	\hbox{Si}\left(
	\frac{\pi}{2}\,\frac{h_S/f_C+2y'/f_L}{\lambda/h_G} 
	\right)
	\right]\;,
\end{eqnarray}
where we indicated with $w_S$ and $w_G$ the full widths of the slit
and the grating aperture, respectively, and with $h_S$ and $h_G$ the
corresponding full heights. The $\hbox{Si}(\theta)$ function satisfies 
the parity condition $\hbox{Si}(-\theta)=-\hbox{Si}(\theta)$, and for 
$\theta>0$ it can be approximated using the relation
\begin{equation} \label{eq:Si1}
\hbox{Si}(\theta)=\frac{\pi}{2}-f(\theta)\cos\theta-g(\theta)\sin\theta\;,
\end{equation}
along with rational approximations of $f(\theta)$ and $g(\theta)$ that are
given for $\theta\ge 1$ \cite{AS64}. For $|\theta|<1$, we
can use the following power expansion (expressed, for computational 
convenience, through Horner's algorithm),
%
%
\begin{equation} \label{eq:Si2}
\hbox{Si}(\theta)\approx \theta\biggl(1 
	- \frac{\theta^2}{6} \biggl(\frac{1}{3} 
	- \frac{\theta^2}{20} \biggl(\frac{1}{5} 
	- \frac{\theta^2}{294} \biggr)\biggr)\biggr)\;,
\end{equation}
which guarantees a precision better than $\sim 3\times 10^{-7}$. (This
can be seen by considering the next order in the power expansion
(\ref{eq:Si2}), which is bounded by 
$1/(6\times 20\times 42\times 648)$; see \cite{AS64}.) For an
infinitely long slit ($h_S\gg|y|=|y'|f_C/f_L$), taking into account 
that $\hbox{Si}(\theta)\to\pi/2$ in the limit of $\theta\to\infty$,
we find more simply
\begin{equation} \label{eq:field1}
U_G(x',y')\to U_G(x')
	=\frac{1}{\pi} \left[
	\hbox{Si}\left(
	\frac{\pi}{2}\,\frac{w_S/f_C-2x'/f_L}{\lambda/w_G} 
	\right) +
	\hbox{Si}\left(
	\frac{\pi}{2}\,\frac{w_S/f_C+2x'/f_L}{\lambda/w_G} 
	\right)
	\right]\;.
\end{equation}
This last result is in agreement with previous theoretical studies of 
the problem of coherent illumination of a spectrograph slit and of 
the determination of
the corresponding LSF \cite{VC30,Mi67}. The advantage of the operatorial
approach adopted here is that the characteristics of the spectrograph are 
all contained in the OTO of Eqs.~(\ref{eq:eq4a}) and (\ref{eq:eq4b}), 
regardless of the form of the input field. This makes for a very 
compact derivation of the various results, and also improves the 
traceability of the various hypotheses and approximations involved.

The description of the spectrograph as a combination of linear operators 
also greatly facilitates the programming of numerical codes for the diffraction 
analysis of spectrographs.
In particular, computational tests using a two-dimensional numerical 
implementation of 
the general OTO of Eq.~(\ref{eq:genOTO}) demonstrate our former argument 
that \emph{the FWHM of the LSF of the spectrograph is principally
affected by the smallest of the limiting apertures in the collimator}, 
regardless of whether this aperture corresponds to the collimator lens or 
the grating (or the camera lens), thus supporting the general applicability 
of Eq.~(\ref{eq:field1}), as soon as we identify $w_G$ with the width of 
such an aperture.
That same code has also been used to validate the model of spectral
resolution presented in Sect.~\ref{sec:resolution}.

\begin{figure}[!t]
\centering
\includegraphics[width=.495\hsize]{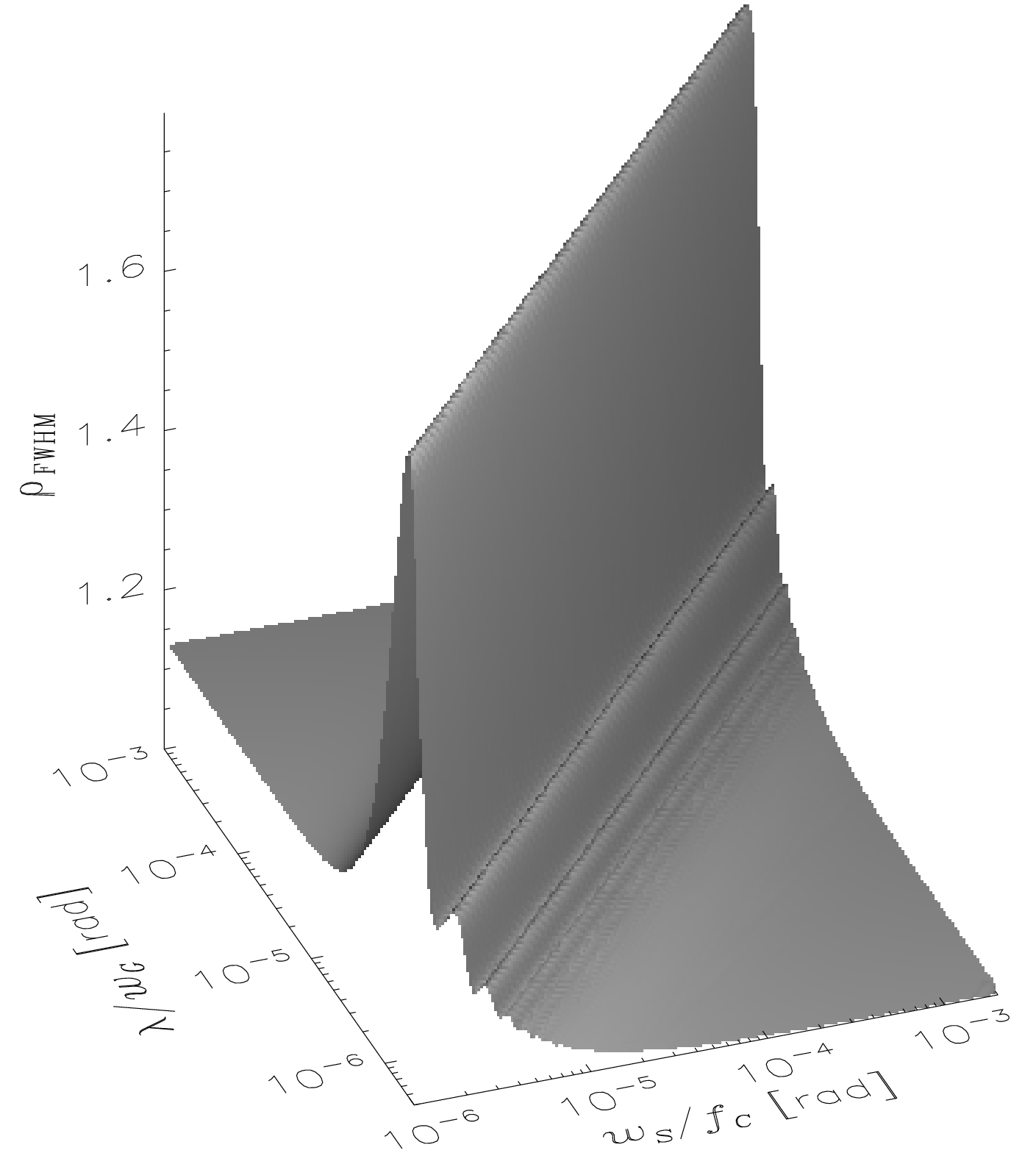}
\includegraphics[width=.495\hsize]{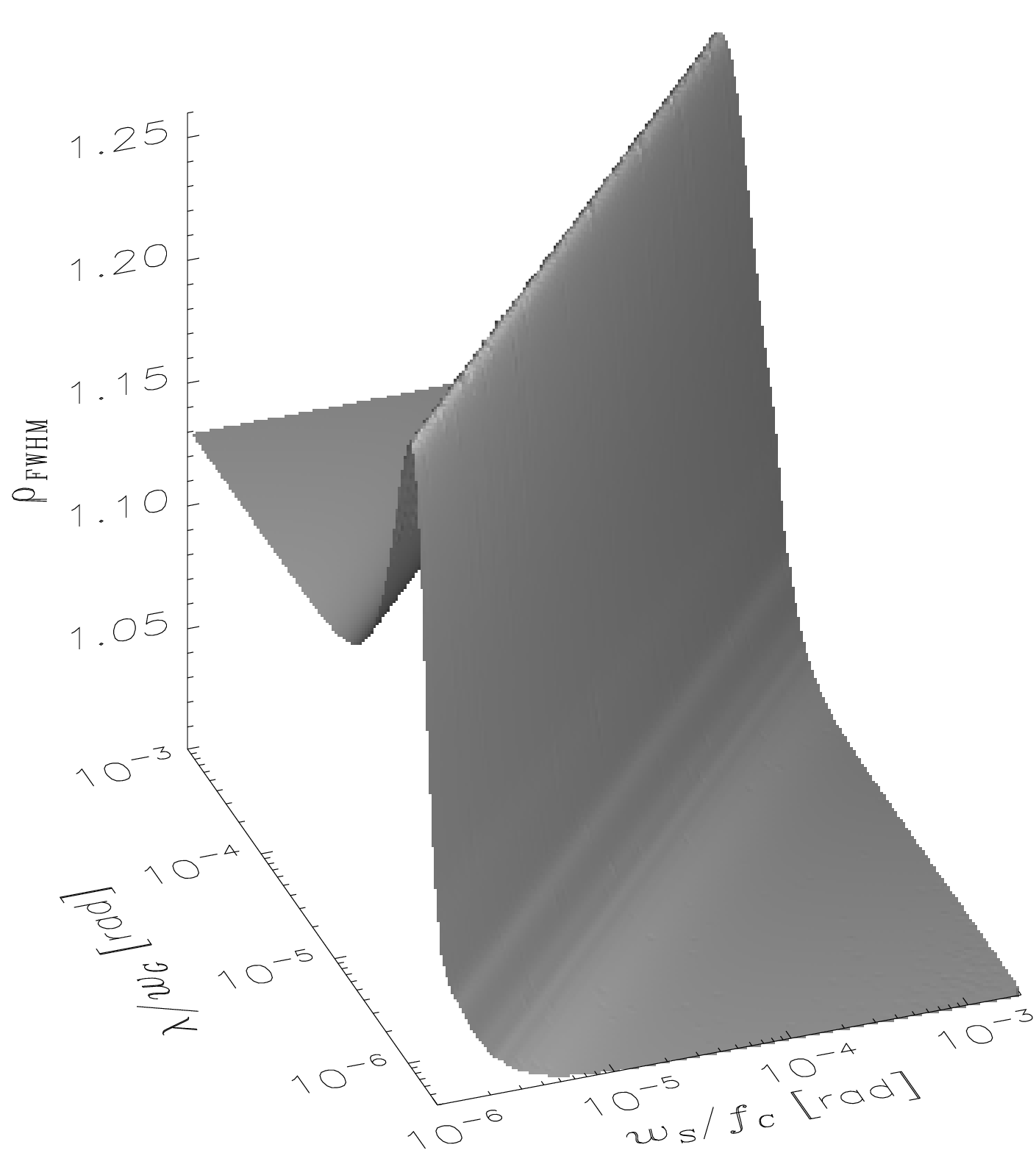}
\caption{\label{fig:FWHMerr}
The ratio $\rho_\mathrm{FWHM}$ between the RSS estimate of the FWHM of 
the spectrograph LSF and the exact value as calculated with the model 
presented in this work, plotted as a function of the two contributions 
$\gamma_S=w_S/f_C$ and $\gamma_G=\lambda/w_G$: 
(left) case of coherent illumination; (right) case of incoherent 
illumination.}
\end{figure}

\begin{figure}[!ht]
\centering
\includegraphics[width=.7\hsize]{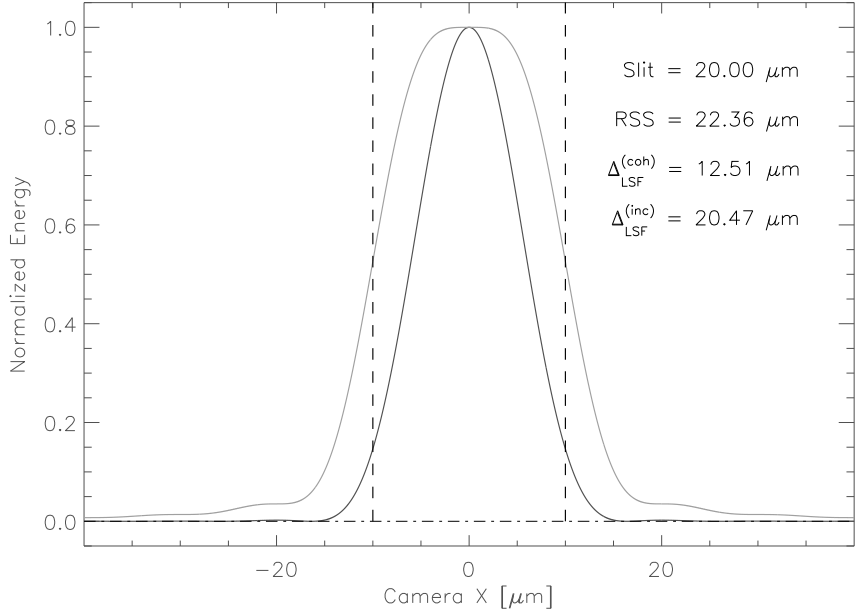}
\caption{\label{fig:FWHMprofile}
Cross-section of the energy distribution at the detector
(normalized to unit peak), from a fully illuminated slit, for a typical 
spectrograph configuration
($\lambda=1\,\rm\mu m$, $w_S=50\,\rm\mu m$, $f_C=2.5$\,m, $w_G=0.1$\,m, 
$f_L=1$\,m). The black (gray) curve is for the case of coherent
(incoherent) illumination. The figure also gives the projected geometric 
slit width, and the RSS estimate of the FWHM.}
\end{figure}

The simple analytic expression for the slit image at the detector given 
by Eq.~(\ref{eq:field1}) provides a convenient and 
accurate method to estimate the \emph{normalized} FWHM of the instrument 
profile, $\Delta x'/f_L$, for the purpose of computing the resolution of 
a spectrograph. We note that this FWHM is a function of only two 
dimensionless parameters, 
$\gamma_S=w_S/f_C$ and $\gamma_G=\lambda/w_G$, the former 
representing the angular 
spread of the slit width as seen by the collimator, while the latter 
corresponds to the FWHM of the 
diffraction profile corresponding to the aperture of the grating (or other 
limiting aperture in the spectrograph).

The left panel of Fig.~\ref{fig:FWHMerr} shows the ratio between the RSS 
estimate and the exact value of the FWHM of the instrument profile 
$|U_G(x')|^2$, as a function of the two parameters $\gamma_S$ and 
$\gamma_G$. The parameter ranges encompass a large set of 
spectrograph configurations. The range
of $\gamma_S$ spans from, say, a $5\,\rm\mu m$ slit aperture
with a collimator of 5\,m focal length, to a 0.5\,mm slit
aperture with a collimator of only 25\,cm focal length. The range
of $\gamma_G$ spans from, say, a wavelength of 200\,nm with a
grating aperture of 50\,cm, to a wavelength of $25\,\rm\mu m$ with a
grating aperture of only 2.5\,cm. This plot demonstrates the error that 
can be expected when the FWHM of the instrument profile is approximated 
by the RSS of the $\gamma_S$ and $\gamma_G$ contributions.
We also see that this approximation becomes accurate for large slit widths 
(relative to the focal length of the collimator) and/or
small wavelengths (relative to the size of the grating), i.e., when
diffraction effects are comparatively unimportant. On the other hand,
for small slit apertures and/or large wavelengths the RSS approximation
consistently overestimates (by about 13\%) the true FWHM. The maximum
error that is incurred by using the RSS approximation in intermediate 
cases is an overestimation of the true FWHM by about 80\%, in the case
of coherent illumination.

The black curve in Figure~\ref{fig:FWHMprofile} shows the LSF at the 
detector (normalized to unit peak), as calculated through 
Eq.~(\ref{eq:field1}), 
for a typical spectrograph configuration corresponding to 
$\gamma_S=2\times 10^{-5}$\,rad and $\gamma_G=10^{-5}$\,rad
(see caption). The location of this configuration in the domain 
of Fig.~\ref{fig:FWHMerr} falls in the region where the departure
between the true FWHM and its RSS estimate is the largest for coherent 
illumination. The
figure also reports the geometrically projected slit (in this case,
corresponding to a spectrograph magnification $f_L/f_C=0.4$).

In the application of Eq.~(\ref{eq:field1}) to realistic grating 
spectrographs, the two parameters $\gamma_S$ and $\gamma_G$ 
must in general be multiplied by the \emph{anamorphic magnification}, 
$r=\cos\alpha/\cos\beta$, where $\alpha$ and
$\beta$ are respectively the incident and diffraction angles of the
radiation measured from the grating normal. Alternatively, if 
$\Delta x'$ is the linear FWHM of the profile of Eq.~(\ref{eq:field1}), 
the actual FWHM of the LSF at the detector plane, $\Delta x'_r$, is 
determined by the condition $\Delta x'=\Delta x'_r/r$. We note in 
particular that
$r\,\gamma_G=(\lambda/\cos\beta)/(w_G/\cos\alpha)
	\equiv\lambda/(L\cos\beta)$, where $L$ is the effective width
of the illuminated grating area (rather than its projected width, $w_G$). 
As expected, this quantity corresponds to the FWHM of the grating
profile function (see, e.g., Eq.~[3-6] in \cite{Gr76}).

%
%

\subsection{The Case of Incoherent Illumination}

Because Eq.~(\ref{eq:eq6b}) has the form of a convolution product
of the input field (weighted by the slit aperture) with the 
translationally invariant kernel
\begin{equation} \label{eq:PSF}
h(\bm{x'},\bm{x})=
	\tilde{t}_G\biggl(\frac{1}{\lambda f_L}
	\left(\bm{x'}-\bm{x}\right)\biggr)\;,
\end{equation}
it follows in particular that $h(\bm{x'},\bm{x})$ represents the 
\emph{impulse response} of the linear system describing the 
spectrograph, under the current assumption that the limiting aperture 
is provided by the grating. This enables a straightforward description
of the behavior of the spectrograph also in the case of fully incoherent
illumination, since in such case \cite{Go96} 
\begin{equation}
I_G(\bm{x'})=A^{-1}\int d\bm{x}\;|h(\bm{x'},\bm{x})|^2\,
t_S(\bm{x})\,I(\bm{x})\;,
\end{equation}
where $I(\bm{x})$ and $I_G(\bm{x'})$ are the field \emph{intensity}
at the entrance slit and at the detector, respectively, and $A$ is a
normalization constant corresponding to the area of the aperture. Hence, 
in the case of incoherent illumination, the result analog to 
Eq.~(\ref{eq:eq6b}) is
\begin{equation}
I_G(\bm{x'}) \label{eq:incoherent}
=\frac{k}{w_G h_G} \int d\bm{\chi}\;
	\biggl|\tilde{t}_G\biggl(\frac{1}{\lambda f_L}
	\left(\bm{x'}-\bm{\chi}\right)\biggr)\biggr|^2\,
	t_S\biggl(-\frac{f_C}{f_L}\bm{\chi}\biggr)
	I\biggl(-\frac{f_C}{f_L}\bm{\chi}\biggr)\;.
\end{equation}
Remarkably, also this expression can be formally integrated in terms of 
the sine integral function. For simplicity of notation, we introduce 
the following dimensionless quantities,
\begin{equation} \label{eq:theta_def}
\theta_x^{\pm}=\frac{\pi}{2}\,\frac{w_S/f_C \pm 2x'/f_L}{\lambda/w_G}\;,
\qquad
\theta_y^{\pm}=\frac{\pi}{2}\,\frac{h_S/f_C \pm 2y'/f_L}{\lambda/h_G}\;.
\end{equation}
We then find explicitly
\begin{eqnarray} \label{eq:field_inc}
I_G(x',y') = \frac{1}{\pi^2}
&&\biggl[
	\hbox{Si}\bigl(2\theta_x^-\bigr) + 
	\hbox{Si}\bigl(2\theta_x^+\bigr)
- \frac{\sin^2\theta_x^-}{\theta_x^-}
- \frac{\sin^2\theta_x^+}{\theta_x^+}
\biggr] \nonumber \\
\times
&&\biggl[
	\hbox{Si}\bigl(2\theta_y^-\bigr) + 
	\hbox{Si}\bigl(2\theta_y^+\bigr)
- \frac{\sin^2\theta_y^-}{\theta_y^-}
- \frac{\sin^2\theta_y^+}{\theta_y^+}
\biggr]\;.
\end{eqnarray}
This expression must be compared with the result of Eq.~(\ref{eq:field}),
\begin{equation}
U_G(x',y') = \frac{1}{\pi^2}
\Bigl[
	\hbox{Si}\bigl(\theta_x^-\bigr) + 
	\hbox{Si}\bigl(\theta_x^+\bigr)
\Bigr]
\Bigl[
	\hbox{Si}\bigl(\theta_y^-\bigr) + 
	\hbox{Si}\bigl(\theta_y^+\bigr)
\Bigr]\;,
\end{equation}
which is instead valid in the case of coherent illumination. Similarly,
in the limit of infinitely long slit, $\theta_y^\pm\to\infty$, and 
Eq.~(\ref{eq:field_inc}) reduces to
\begin{equation} \label{eq:field1_inc}
I_G(x',y')\to I_G(x')
	=\frac{1}{\pi} \biggl[
	\hbox{Si}\bigl(2\theta_x^-\bigr) + 
	\hbox{Si}\bigl(2\theta_x^+\bigr)
- \frac{\sin^2\theta_x^-}{\theta_x^-}
- \frac{\sin^2\theta_x^+}{\theta_x^+}
\biggr]\;.
\end{equation}
This result corresponds to the one originally derived by \cite{Wa97}
(see also \cite{VC30,Mi67}). Once again, in the application of
Eq.~(\ref{eq:field1_inc}) to realistic grating spectrographs, the 
quantities $\gamma_S=w_S/f_C$ and $\gamma_G=\lambda/w_G$ appearing in
the definition of $\theta_x^\pm$, Eq.~(\ref{eq:theta_def}), must be 
multiplied by the proper anamorphic magnification $r$.

The right panel of Fig.~\ref{fig:FWHMerr} shows the ratio between the RSS 
estimate and the exact value of the FWHM of the instrument profile 
$I_G(x')$, as a function of the two parameters $\gamma_S$ and 
$\gamma_G$. We note that
using the RSS estimate of the FWHM in the incoherent case can lead to 
an underestimation of the spectral resolution of the instrument by as
much as 25\%.
Correspondingly, the gray curve of Fig.~\ref{fig:FWHMprofile} shows the 
LSF at the detector (normalized to unit peak) in the case of incoherent 
illumination, for the spectrograph configuration listed in the figure's
caption.

It is worth noting that \emph{the FWHMs of the spectrograph LSF for the 
two cases of coherent and incoherent illumination tend to the same value 
in both regimes $\gamma_S\gg\gamma_G$ and $\gamma_S\ll\gamma_G$.} For 
intermediate regimes, the FWHM in the coherent case is \emph{always 
smaller} than the FWHM in the incoherent case, the latter being larger
by as much as 40\%. 

\section{Spectral Resolution}
\label{sec:resolution}

The results of Eqs.~(\ref{eq:field1}) and (\ref{eq:field1_inc}) do not 
take into account the 
sampling of the LSF by the pixels of the detector. Therefore the 
FWHM of the instrument profile that is derived from those equations, and 
which we hereafter indicate with $\Delta_{\rm LSF}$, cannot yet be 
associated with the effective spectral resolution of the instrument. Let 
us indicate with $\zeta$ the ratio of $\Delta_{\rm LSF}$ to the
pixel size $\delta_{\rm cam}$ of the detector expressed in the 
same units, 
\begin{equation}	\label{eq:zeta}
\zeta=\frac{\Delta_{\rm LSF}}{\delta_{\rm cam}}\;.
\end{equation}
We want to be able to model the minimum 
resolvable spectral interval 
$\Delta\lambda$ as a function of $\zeta$, so that we can estimate the 
spectral resolution of the instrument as
\begin{equation}	\label{eq:spectral_resol}
R(\lambda;\zeta)=\frac{\lambda}{\Delta\lambda(\zeta)}\;.
\end{equation}
In order to do this, it is convenient to express $\Delta_{\rm LSF}$ 
and $\delta_{\rm cam}$ in wavelength units. If $\delta\beta$ is some 
angular interval in the diffracted beam, the corresponding wavelength 
interval for a given spectrograph configuration $(\alpha,\beta)$ is 
\begin{equation} \label{eq:disper}
\delta\lambda=\frac{d\lambda}{d\beta}\,\delta\beta
	=\frac{\lambda\cos\beta}{\sin\alpha+\sin\beta}\,\delta\beta\;,
\end{equation}
where in the last equivalence we used the expression of the angular
dispersion $d\beta/d\lambda$, which is derived from 
the grating 
equation. 
In the following, we assume that both quantities 
$\Delta_{\rm LSF}$ and $\delta_{\rm cam}$ have been converted to 
wavelength units through the transformation (\ref{eq:disper}). Of course, 
if those quantities were originally given in linear units, they must 
first be transformed to angular units through division by $f_L$.

Knowledge of the detailed physical behavior of the function 
$\Delta\lambda(\zeta)$ in Eq.~(\ref{eq:spectral_resol}) is often not
necessary, if we are only interested in approximately estimating 
the spectral resolution of the instrument (say, within 10\%). This
is the case, for example, in many spectrographic applications to the 
remote sensing of astrophysical plasmas. On the other hand, the function 
$\Delta\lambda(\zeta)$ must always satisfy the two following limit conditions: 
1) for very small $\zeta$, the LSF of the instrument is completely 
undersampled, and the spectral resolution is thus determined by the 
detector's pixel size, so that
\begin{equation} \label{eq:lim1}
\lim_{\zeta\to 0}\Delta\lambda(\zeta)\approx 2\delta_{\rm cam}\;,
\end{equation}
where the (approximate) factor 2 enters because of the Nyquist criterion; 
2) for very large $\zeta$, the LSF is
fully resolved by the detector, and therefore the spectrograph 
resolution is simply determined by $\Delta_{\rm LSF}$, that is,
\begin{equation} \label{eq:lim2}
\lim_{\zeta\to\infty}\Delta\lambda(\zeta)=\Delta_{\rm LSF}\;.
\end{equation}
We also note that the condition (\ref{eq:lim2}) must be approximately satisfied 
already for relatively small values of $\zeta$, practically as soon as the
LSF is critically sampled by the detector, i.e., $\zeta\gtrsim 2$, or
$\Delta_{\rm LSF}\gtrsim 2\delta_{\rm cam}$. On the other hand, for 
$\zeta\approx1$, we have the so-called ``pixel matching'' condition for 
a spectrograph, in which $\Delta_{\rm LSF}\approx\delta_{\rm cam}$. 
Under this condition, the minimum resolved spectral interval is evidently 
$\Delta\lambda\approx 2\Delta_{\rm LSF}\approx 2\delta_{\rm cam}$.

\begin{figure}[t!]
\centering
\includegraphics[width=.6\hsize]{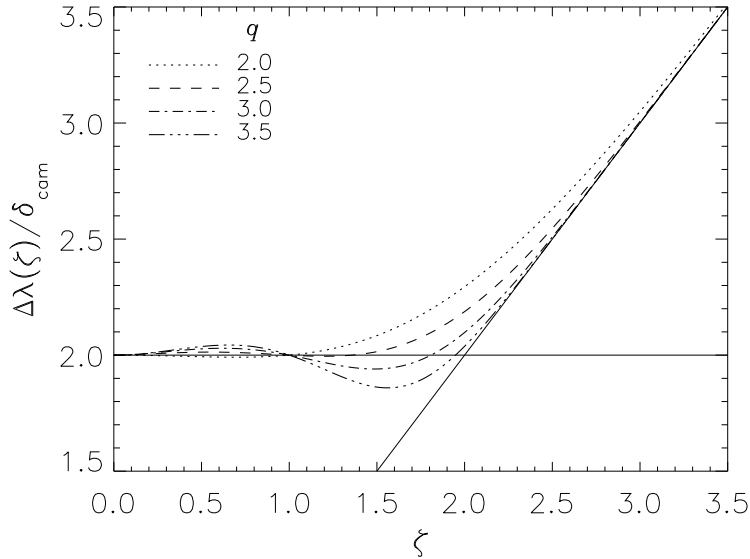}
\caption{\label{fig:resolution}
Plots of Eq.~(\ref{eq:FWHM}) with $\alpha=\ln(4/3)$, for various values 
of the exponent $q$. See text for details.}
\end{figure}

We can then conclude that the function $\Delta\lambda(\zeta)$ must remain
approximately at the value $2\delta_{\rm cam}$ for $0\le\zeta\lesssim 2$, to 
then rapidly turn into a linear function of $\zeta$ for
$\zeta\gtrsim 2$, so that $\Delta\lambda(\zeta)\sim\Delta_{\rm LSF}$ in that
regime. We thus propose the following functional form for 
$\Delta\lambda(\zeta)$,
\begin{equation} \label{eq:FWHM}
\Delta\lambda(\zeta)=\delta_{\rm cam}\sqrt{\zeta^2+4\exp(-\alpha\zeta^q)}\;,
\end{equation}
where $\alpha$ and $q$ are two positive quantities. We see that
Eq.~(\ref{eq:FWHM})
automatically satisfies the limits (\ref{eq:lim1}) and 
(\ref{eq:lim2}). The value of 
$\alpha$ can be determined by imposing the additional condition that, for 
pixel matching ($\zeta=1$), it also must be 
$\Delta\lambda(1)\approx 2\delta_{\rm cam}$. If, for simplicity, we assume 
this condition to be exact, we find 
\begin{equation} \label{eq:alpha}
\alpha=\ln(4/3)\;.
\end{equation}
The parameter $q$ remains undetermined, and can be chosen so to provide 
a good representation of the qualitative behavior of
$\Delta\lambda(\zeta)$ 
as discussed above. 
Figure~\ref{fig:resolution} shows the graphs of Eq.~(\ref{eq:FWHM}), 
with the condition (\ref{eq:alpha}), for several
values of $q$. We find that $2.5\lesssim q\lesssim3.0$ provides a good 
choice for approximating the expected behavior of 
$\Delta\lambda(\zeta)$ as a function of the sampling ratio. 

\begin{figure}[!ht]
\centering
\includegraphics[width=.7\hsize]{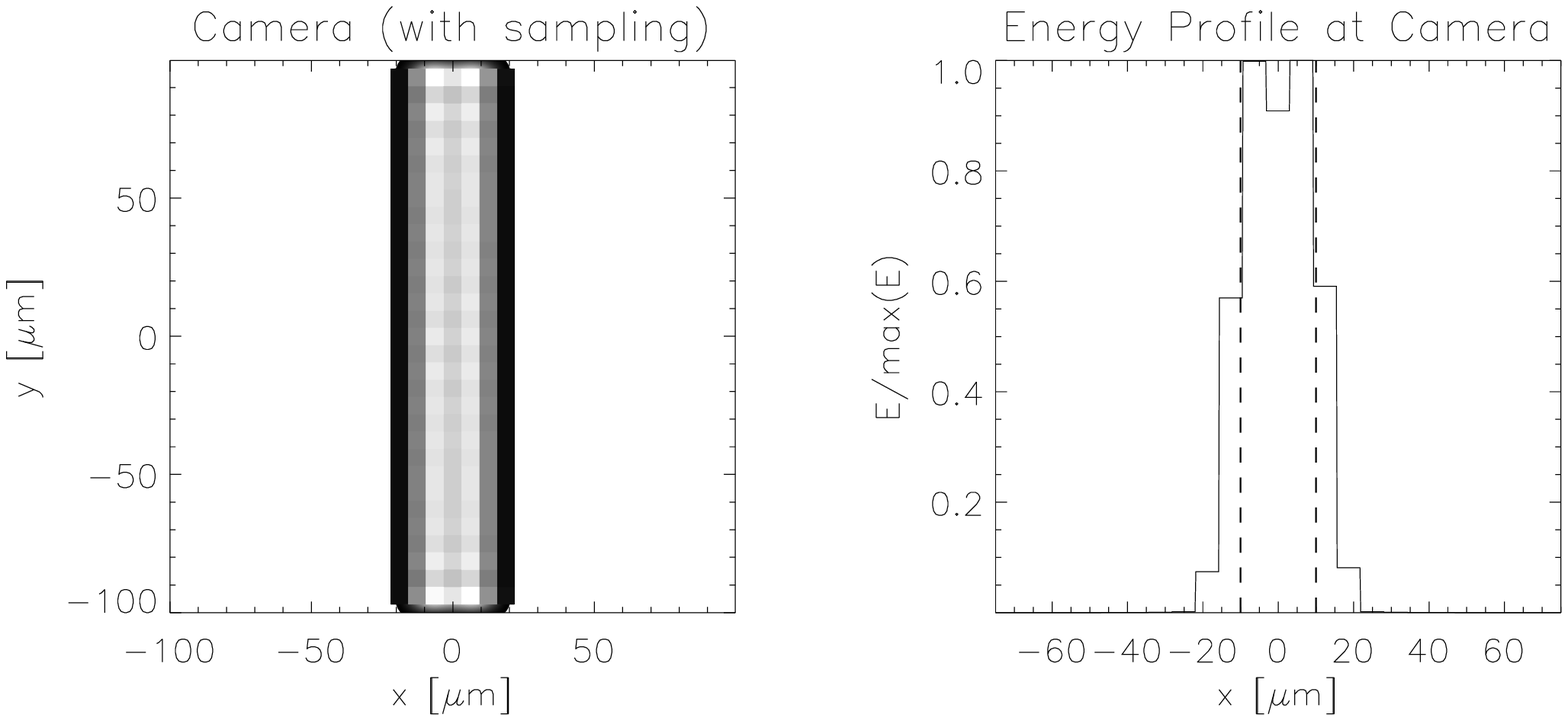}
\includegraphics[width=.7\hsize]{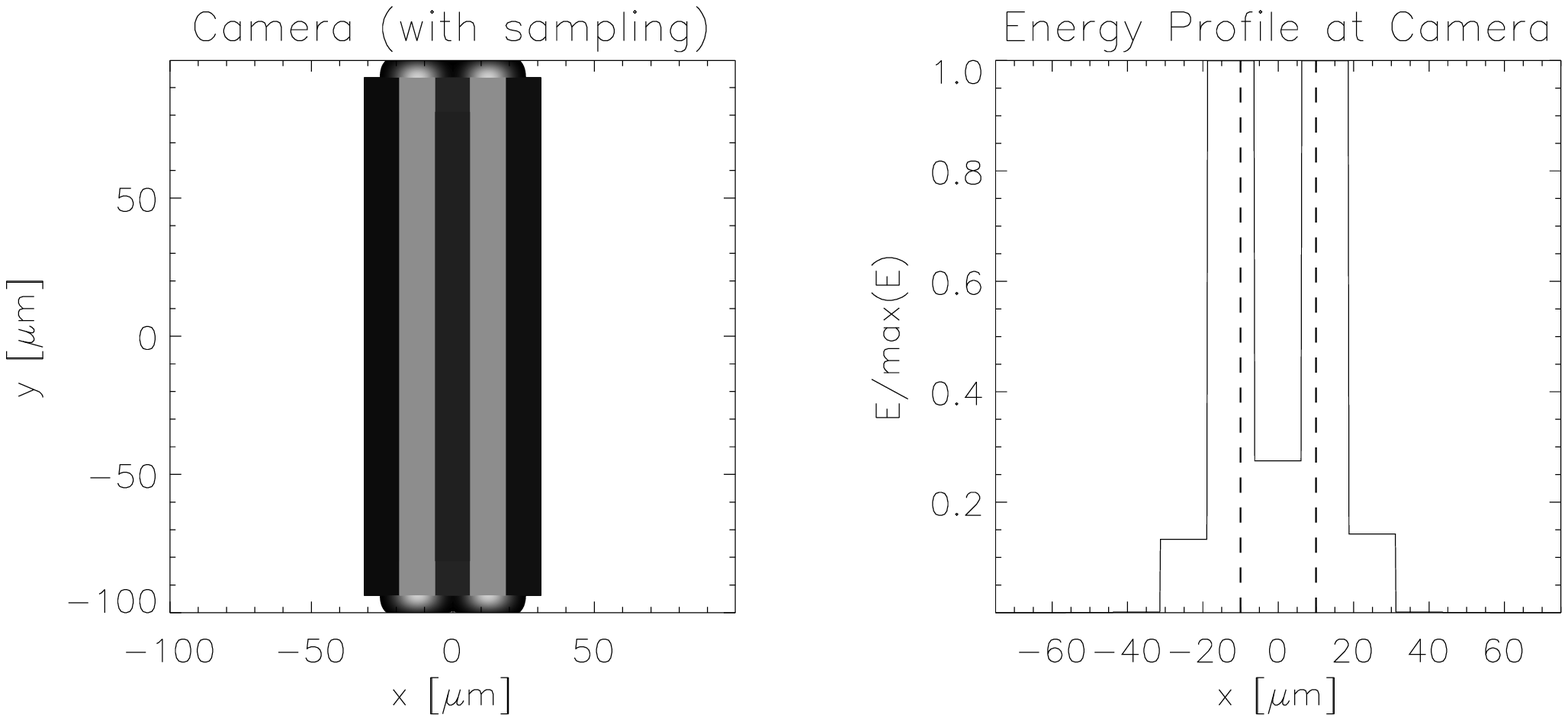}
\includegraphics[width=.7\hsize]{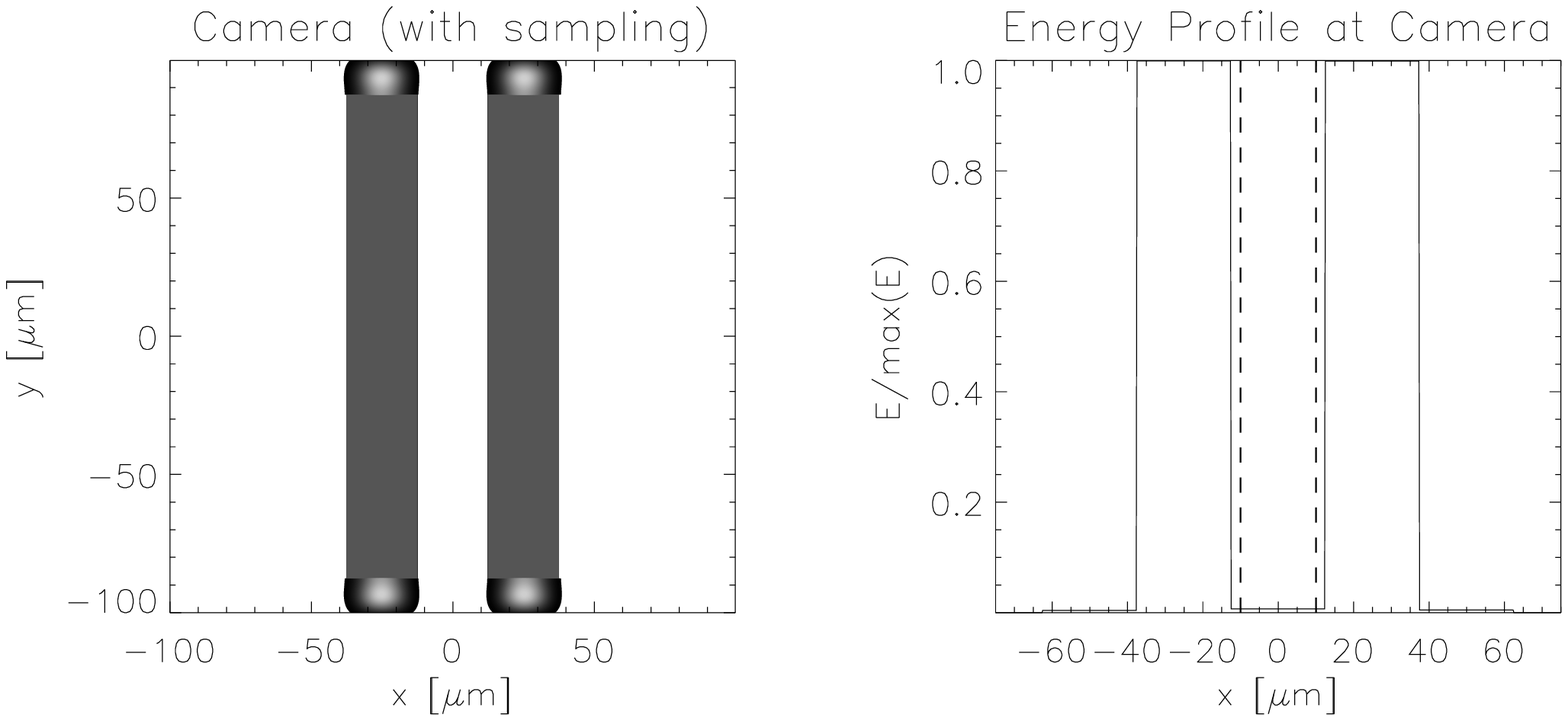}
\caption{\label{fig:test_resol}
Numerical tests of the resolution formula Eq.~(\ref{eq:FWHM}), using the 
same spectrograph configuration of Fig.~\ref{fig:FWHMprofile}, for three
different pixel sizes: 
$6.25\,\rm\mu m$ (top), $12.5\,\rm\mu m$ (center), and $25\,\rm\mu m$
(bottom), corresponding to $\zeta=2.0,1.0,0.5$, respectively. See text
for details. The vertical dashed lines in the panels on the right define 
the geometric projection of the slit (cf.~Fig.~\ref{fig:FWHMprofile}).}
\end{figure}

To summarize, we propose the use of the expressions 
(\ref{eq:spectral_resol}) and (\ref{eq:FWHM}), along with the definition 
(\ref{eq:zeta}), for an improved estimate of the spectral resolution of
a slit-based spectrograph, under various regimes of slit diffraction and 
detector sampling. The FWHM of the instrument profile, 
$\Delta_\mathrm{LSF}$, must be computed numerically through 
Eq.~(\ref{eq:field1}) in the case of coherent illumination, and through
Eq.~(\ref{eq:field1_inc}) in the case of incoherent illumination, after 
due modification to account for the 
anamorphic magnification corresponding to the specific configuration of 
the spectrograph (see discussion at the end of Section~\ref{sec:derivation}). 
We suggest a value between 2.5 and 3.0 for the $q$ exponent of 
$\zeta$ in Eq.~(\ref{eq:FWHM}), since this provides a good qualitative 
agreement with the proposed model of spectral resolution, as it is
apparent from the plots of Figure~\ref{fig:resolution}.

Figure~\ref{fig:test_resol} shows two-dimensional numerical diffraction 
calculations of the instrument profile of a spectrograph sampled by 
a detector with which we tested the validity of Eq.~(\ref{eq:FWHM}). 
For the test, we adopted the same spectrograph configuration, slit 
width, and wavelength as in the example of Fig.~\ref{fig:FWHMprofile}, 
for the case of coherent illumination. We considered three different 
sampling sizes of the detector's pixel, namely, $\delta_{\rm cam}=6.25$, 
12.5, and $25\,\rm\mu m$, corresponding to the cases of $\zeta=2.0$, $1.0$, 
and $0.5$, respectively. Finally, we used our numerical code implementing
the general OTO of Eq.~(\ref{eq:genOTO}) to compute the two-dimensional
LSF and replicate it at the positions on the detector of two neighboring 
wavelengths exactly separated by the spectral resolution interval 
$\Delta\lambda(\zeta)$, as calculated through Eq.~(\ref{eq:FWHM}). 
In practical cases, the resulting pattern would correspond to two 
infinitely sharp emission lines that are barely resolvable by the
spectrograph.
The panels on the left are contour plots of the diffracted energy
at the two neighboring wavelengths. Those plots show the appearance of 
the two infinitely sharp emission lines after being diffracted through 
the spectrograph and sampled by the detector. We displayed the 
\emph{unsampled} diffracted field at the very top and bottom pixels of
the contour plots, so to help visualize how the detector array aligned 
with the LSF for each of the test cases.
The panels on the right show the cross-cut energy
profile (normalized to unit peak amplitude) of the contour plot on the
left, for $y=0$. 
These calculations show that the two neighboring wavelengths are 
either fully or just barely resolved, depending on whether the position 
of the sampling array is more or less optimally aligned with the LSF profile. 
This is indeed what we must expect from a definition of \emph{minimum} 
spectral resolution based on the Nyquist criterion.

We want to conclude this section by comparing our results for the 
resolution of a grating spectrograph with other commonly used
expressions that are found in the literature. We refer particularly 
to the comprehensive treatment of spectrographs presented in \cite{Sc00}. 
Using the assumptions of that work, the spectral resolution becomes
in our notation (cf.\ Eq.~[12.2.3] of \cite{Sc00})
\begin{equation} \label{eq:Sc0}
R=\frac{\sin\alpha+\sin\beta}{\cos\beta}\,\frac{w_G}{r w_S}\,
	\frac{F}{D}\;,
\end{equation}
where $D$ and $F$ are, respectively, the diameter and focal length of 
the telescope feeding the spectrograph. The ``seeing-limited'' (s.l.)
case corresponds to the regime $\gamma_S\gg\gamma_G$. Hence, diffraction 
effects from the slit can be ignored, and it is possible to apply the 
geometric condition $w_G/f_C=D/F$ that corresponds to the conservation 
of the \emph{etendue} \cite{Sc00}. Equation~(\ref{eq:Sc0}) then becomes
\begin{equation} \label{eq:Sc1}
R_\mathrm{s.l.}=\frac{\sin\alpha+\sin\beta}{\cos\beta}\,\frac{f_C}{r w_S}
 =\frac{\sin\alpha+\sin\beta}{\cos\beta}\,\frac{1}{r\,\gamma_S}\;.
\end{equation}
On the other hand, in the ``diffraction-limited'' (d.l.) case, 
$w_S=(\lambda/D)F$, and Eq.~(\ref{eq:Sc0}) becomes
\begin{equation} \label{eq:Sc2}
R_\mathrm{d.l.}=\frac{\sin\alpha+\sin\beta}{\cos\beta}\,
\frac{w_G}{r \lambda}
 =\frac{\sin\alpha+\sin\beta}{\cos\beta}\,\frac{1}{r\,\gamma_G}\;.
\end{equation}

If we ignore the sampling of the LSF by the detector, from 
Eqs.~(\ref{eq:spectral_resol}) and (\ref{eq:disper}) we find instead
that the expression for the spectral resolution has the general form
\begin{equation}
R=\frac{\sin\alpha+\sin\beta}{\cos\beta}\,
	\frac{1}{\Delta_{\rm LSF}}\;,
\end{equation}
\emph{for all regimes} of $\gamma_S$ and $\gamma_G$. 
Thus our expression coincides with one or the other of 
Eqs.~(\ref{eq:Sc1}) and (\ref{eq:Sc2}) only if
$\Delta_{\rm LSF}=r\,\gamma_S$ or
$\Delta_{\rm LSF}=r\,\gamma_G$. 
Our previous analysis has shown that indeed 
$\Delta_{\rm LSF}\to r\,\sqrt{\gamma_S^2+\gamma_G^2}
	\to r\,\gamma_S$
when $\gamma_S\gg\gamma_G$, 
whereas the general expression for $\Delta_{\rm LSF}$ never tends to
$r\,\gamma_G$ (see Fig.~\ref{fig:FWHMerr}) when $\gamma_S\ll\gamma_G$. 
To demonstrate this, we can simply consider the case of coherent
illumination in the regime $\gamma_S\ll\gamma_G$, since we already noted 
that both $|U_G(x')|^2$ and $I_G(x')$ give rise to the same 
$\Delta_{\rm LSF}$ in that regime. If we then let 
$\gamma_S=\epsilon\,\gamma_G$ in Eq.~(\ref{eq:field1}), we find, in the 
limit $\epsilon\ll 1$,
\begin{equation}
|U_G(x')|^2\sim \epsilon^2\,\mathrm{sinc}^2
\left(\frac{x'/f_L}{r\,\gamma_G}\right)\;,
\end{equation}
%
so that $\Delta x'/f_L\approx 0.8859\,r\,\gamma_G$. Therefore, while 
the use of Eq.~(\ref{eq:Sc1}) is appropriate in the slit-dominated regime 
of spectrographic observations, the use of Eq.~(\ref{eq:Sc2}) near the
diffraction limit 
leads to an underestimation (by about 13\%) of the true spectral resolution 
(see Fig.~\ref{fig:FWHMerr}).

\begin{acknowledgments}
The authors are thankful to P.\ Nelson and P.\ Oakley (HAO) for helpful 
discussions during the investigation of this problem, and to P.\ Judge
(also HAO) for a careful reading of the manuscript and valuable suggestions.
\end{acknowledgments}

\section{Appendix: The Operational Calculus of Fourier Optics}

In this Appendix we summarize the main properties of the Fourier 
operational calculus for the analysis of optical systems \cite{NS80,Go96}. As
different authors use slightly different definitions, we decided to
adopt the original presentation of \cite{NS80}.
The four basic operations other than the identity are: 

\begin{enumerate}[leftmargin=*]

\item \emph{Multiplication by a quadratic-phase exponential},
\begin{equation}
\op{Q}{c}\{f(\bm{u})\}=\exp\left(\imag\,\frac{\pi}{\lambda}\,c
|\bm{u}|^2\right)f(\bm{u})\;,
\end{equation}
which satisfies the group properties $\op{Q}{a}\op{Q}{b}=\op{Q}{a+b}$
and $\op{Q}{0}=\mathbf{1}$, 
for any parameters $a$ and $b$, and operand $f(\bm{u})$.

\item \emph{Scaling by a constant},
\begin{equation} \label{eq:scaling}
\op{V}{c}\{f(\bm{u})\}=f(c\bm{u})\;,
\end{equation}
which satisfies the group properties $\op{V}{a}\op{V}{b}=\op{V}{ab}$
and $\op{V}{1}=\mathbf{1}$. We note that, if 
$h(\bm{u})=f(\bm{u})\otimes g(\bm{u})$ (with ``$\otimes$'' an
arbitrary multiplicative operation), then Eq.~(\ref{eq:scaling}) 
applied to $h(\bm{u})$ implies the distributive property
\begin{equation} \label{eq:scaling2}
\op{V}{c}\{f(\bm{u})\otimes g(\bm{u})\}
	= \op{V}{c}\{f(\bm{u})\}\otimes\op{V}{c}\{g(\bm{u})\}\;.
\end{equation}

\item \emph{Direct and inverse Fourier Transform (FT)},
\begin{equation}
\mathscr{F}^{\pm 1}\{f(\bm{u})\}=\int\limits_{-\infty}^{+\infty}
	d\bm{u}\,
	\exp(\mp\,\imag\,2\pi\,\bm{u'}\cdot\bm{u})\,f(\bm{u})\;,
\end{equation}
with $\mathscr{F}\mathscr{F}^{-1}=\mathscr{F}^{-1}\mathscr{F}=\mathbf{1}$.
Obviously, the conjugate variables $\bm{u}$ and $\bm{u'}$ must satisfy
the dimensional relation $[u']=[u]^{-1}$.

\item \emph{Free-space propagation} of a field $U(\bm{x})$ through a 
distance $d$,
\begin{equation} \label{eq:freeprop}
\op{R}{d}\{U(\bm{x})\}=\frac{\exp(\imag\,2\pi\,d/\lambda)}{\imag\lambda d}
	\int\limits_{-\infty}^{+\infty}d\bm{x}\,
	\exp\left(\imag\,\frac{\pi}{\lambda d}\,|\bm{x'}-\bm{x}|^2\right)
	U(\bm{x})\;,
\end{equation}
which satisfies the group properties $\op{R}{a}\op{R}{b}=\op{R}{a+b}$
and $\op{R}{0}=\mathbf{1}$. These properties are not self-evident, but
they easily follow as a corollary of the equivalence
\begin{equation}
\op{R}{d}=\frac{\exp(\imag\,2\pi\,d/\lambda)}{\imag}\,
	\mathscr{F}^{-1}\op{Q}{-\lambda^2 d}\mathscr{F}\;,
\end{equation}
which is derived from applying the convolution theorem
to the expression of Eq.~(\ref{eq:freeprop}). Another useful expression of the 
free-space propagation operator is the following,
\begin{equation}
\op{R}{d}=\frac{\exp(\imag\,2\pi\,d/\lambda)}{\imag\lambda d}\,
\op{Q}{\frac{1}{d}} \op{V}{\frac{1}{\lambda d}}
	\mathscr{F} \op{Q}{\frac{1}{d}}\;.
\end{equation}
\end{enumerate}

The operators listed above satisfy several commutation relations that 
are useful for the manipulation of chains of operators representing
optical systems. Some of these relations, which have been used in 
these notes, are the following:
\begin{eqnarray}
\op{V}{t}\mathscr{F}
&=&\mathscr{F}\op{V}{1/t}\;, \\
\op{V}{t}\op{Q}{c}
&=&\op{Q}{t^2 c}\op{V}{t}\;, \\
\mathscr{F}^2
&=&\op{V}{-1}\;.
\end{eqnarray}

\end{document}